\documentclass[aps,floats,showpacs,epsf,superscriptaddress,twocolumn,amsmath]{revtex4}
\DeclareMathAlphabet{\mathpzc}{OT1}{pzc}{m}{it} 
\usepackage{amsfonts, relsize, color}
\usepackage{graphics}
\usepackage{graphicx}
\usepackage{subfigure}
\usepackage{hyperref}

\begin{document}

\title{Dirty Weyl semimetals: Stability, phase transition and quantum criticality}

\author{Soumya Bera}
\affiliation{Max-Planck-Institut f\"ur Physik Komplexer Systeme, 01187 Dresden, Germany}

\author{Jay D. Sau}
\affiliation{Condensed Matter Theory Center, Department of Physics, University of Maryland, College Park, MD 20742, USA}

\author{Bitan Roy}
\affiliation{Condensed Matter Theory Center, Department of Physics, University of Maryland, College Park, MD 20742, USA}

\date{\today}

\begin{abstract}
We study the stability of three-dimensional incompressible Weyl semimetals in the presence of random quenched charge impurities. Combining numerical analysis and scaling theory we show that in the presence of sufficiently weak randomness (i) Weyl semimetal remains stable, while (ii) double-Weyl semimetal gives rise to compressible diffusive metal where the mean density of states at zero energy is finite. At stronger disorder, Weyl semimetal undergoes a quantum phase transition and enter into a metallic phase. Mean density of states at zero energy serves as the order parameter and displays single-parameter scaling across such disorder driven quantum phase transition. We numerically determine various exponents at the critical point, which appear to be insensitive to the number of Weyl pairs. We also extract the extent of the quantum critical regime in disordered Weyl semimetal and the phase diagram of dirty double Weyl semimetal at finite energies.
\end{abstract}


\maketitle

\vspace{10pt}

\emph{Introduction}: Over the span of last few years the horizon of topological phases of matter has been extended beyond the gapped states~\cite{TI-review-1, TI-review-2} and now includes various nodal (gapless) systems as well~\cite{nagaosa, furusaki}. Three dimensional Weyl semimetal (WSM) is the prime example of such non-insulating systems, which is constituted by so called Weyl nodes that act as source (monopole) and sink (anti-monopoles) for \emph{Berry flux} in the reciprocal space, thus always appear in pairs~\cite{nielsen}. As the hallmark signature of a topologically nontrivial phase, WSM accommodates gapless (chiral) surface states that can give rise to peculiar electro-magnetic responses, such as, anomalous Hall and chiral-magnetic effects~\cite{burkov-review}. While the gapped topological phases are expected to be robust against sufficiently weak randomness, stability of their gapless counterpart against disorder demands careful investigation and constitutes the central theme of this Rapid Communication.

Recent time has witnessed the discovery of WSMs in a number of noncentrosymmetric and magnetic semiconductors~\cite{taas-1, tasas-2, taas-3,nbas-1, tap-1, nbp-1, nbp-2, tas, borisenko, chiorescu}. Various other proposals for WSMs, for example, include anti-ferromagnetically~\cite{iridates} or spin-ice~\cite{iridates-ice} ordered pyrochlore iridates, multilayer configuration of topological and regular insulators~\cite{balents-burkov, burkov-zyunin}, magnetically doped topological insulators~\cite{qi-liu}. The monopole charge of Weyl nodes in these systems is $\pm 1$. Nevertheless, HgCr$_2$Se$_4$~\cite{DWS-1}, SrSi$_2$~\cite{DWS-2} are expected to host Weyl nodes with monopole charge $\pm 2$, dubbed as the double-WSM. The topological invariant and enclosed Berry flux in double-WSM is twice that in a WSM, and consequently the one-dimensional chiral surface states in the former system possess a two-fold degeneracy. Although electro-magnetic responses in Weyl materials are reasonably well understood~\cite{burkov-review}, stability of incompressible topological semimetals in the presence of quenched randomness is yet to be explored and settled. This is the quest that has recently culminated in a surge of analytical~\cite{fradkin, shindou-murakami, ominato-koshino, goswami-chakravarty, radzihovsky-gurarie-1, nandkishore, arovas, skinner, roy-dassarma, altland, juricic} and numerical~\cite{imura,herbut-disorder, brouwer-1, pixley-1, brouwer-2, pixley-2, Ohtsuki, chen-Song, hughes, pixley-rareregion} works and we pursue it here for WSM and double-WSM, using numerical and analytical methods.

\begin{figure}[htb]
\centering
\includegraphics[width=8.75cm, height=3.5cm]{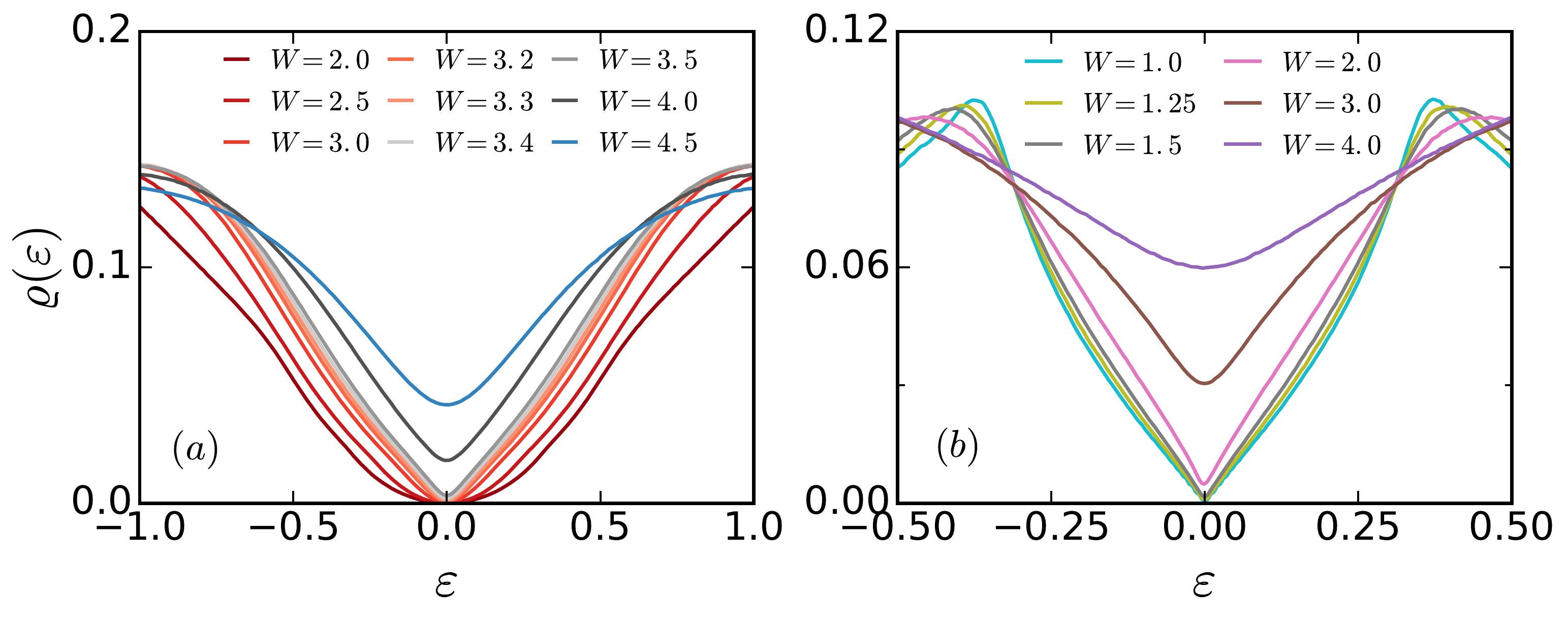}
\caption{(Color online) MDOS in (a) WSM ($N_W=1$) and (b) double-WSM. WSM remains stable up to $W_c=3.3 \pm 0.1$, beyond which the mean DOS at $\varepsilon=0$, $\varrho(0)$ is finite, and the system becomes a CDM. The double-WSM visibly turns into a CDM for weak enough disorder $W \gtrsim 1.0$.}
\label{Fig-1}
\end{figure}

We address the stability of these two systems against random quenched charge impurities by analyzing the mean density of states~(MDOS) at the zero energy, where non-degenerate valence and conduction band touch each other. Our central results are: (i) For sufficiently weak disorder, while WSMs remain stable, double-WSM undergoes a \emph{BCS-like weak coupling instability} toward the formation of a compressible diffusive metal (CDM), where the MDOS at zero energy is finite [see Fig.~\ref{Fig-1}]. (ii) WSMs undergo a disorder driven quantum phase transition (QPT), beyond which the system becomes a CDM. (iii) Across the WSM-CDM transition MDOS display single-parameter scaling and within our numerical accuracy the critical exponents at such \emph{itinerant} quantum critical point (QCP) appear to be insensitive to the number of Weyl pairs ($N_W$) (see Table~\ref{table-1}).

\begin{figure}[htb] 
\includegraphics[width=8.75cm, height=3.5cm]{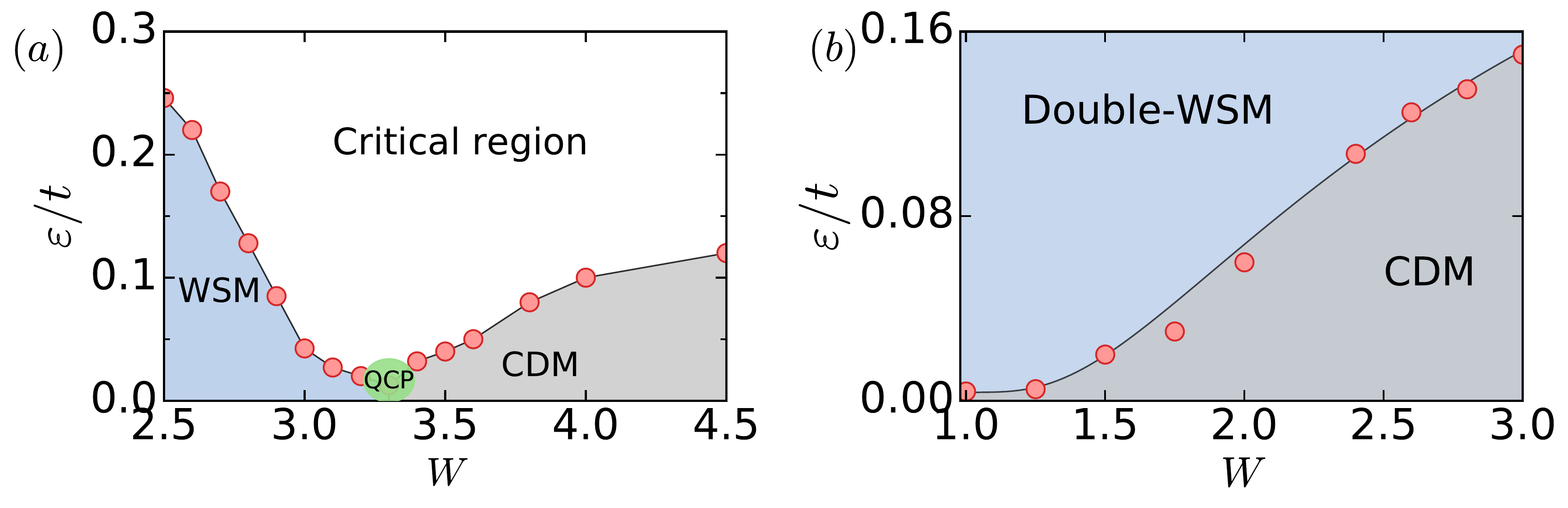}
\caption{(Color online) Finite energy phase diagram of a dirty (a) WSM ($N_W=1$) and (b) double-WSM, for $L=220$. }\label{Fig-4} 
\end{figure}

\emph{Model}: A paradigmatic two-band toy model
\begin{equation}\label{Hgen}
H_W= \sum_{\mathbf{k}} \Psi^\dagger_{\mathbf{k}} \left[ N_1(\mathbf{k}) \sigma_1 + N_2(\mathbf{k}) \sigma_2 + N_3(\mathbf{k}) \sigma_3 \right] \Psi_{\mathbf{k}},
\end{equation}
can describe different members of the Weyl family, where $\boldsymbol \sigma$ are standard Pauli matrices. Fermionic annihilation operators $c_{s, \mathbf{k}}$ with spin-projections $s=\uparrow,\downarrow$ and wave vector $\mathbf{k}$, constitute the two component spinor $\Psi^\top_{\mathbf{k}}=\left( c_{\uparrow, \mathbf{k}}, c_{\downarrow, \mathbf{k}}\right)$. A WSM is found upon choosing $N_j(\mathbf{k})=t \sin(k_j a)$ for $j=1,2$, $N_3(\mathbf{k})= N^{1}_3(\mathbf{k}) + N^{2}_3(\mathbf{k})$, where $2 N^{1}_3(\mathbf{k})=t\cos(k_3 a)$ and $a$ is the lattice spacing. In this model, the number of Weyl pairs ($N_W$) can be tuned efficiently by the \emph{Wilson mass} $N^{2}_3(\mathbf{k})= t'[b-\cos(k_1 a) -\cos(k_2 a)]$. A double-WSM can be constructed by taking $N_1(\mathbf{k})=t_1 [\sin(k_1 a) -\sin(k_2 a)]$, $N_2(\mathbf{k})=t_1 \cos(k_1 a) \cos(k_2 a)$, and $N^{2}_3(\mathbf{k})=t^\prime[2-\sin(k_x a)-\sin(k_y a)]$, while keeping $N^{1}_3(\mathbf{k})$ unaltered. We implement these tight-binding models on a cubic lattice with periodic boundary in each direction~\cite{supplementary}.

\emph{Disorder}: The quintessential properties of dirty Weyl systems can be established from their effective low energy theory in the close vicinity of the Weyl points. The low energy Hamiltonians for WSM and double-WSM are 
\begin{eqnarray}
H_1&=&\Psi^\dagger_\tau \left[ -i v \left( \sigma_1 \partial_1 + \sigma_2 \partial_2 + \tau \sigma_3 \partial_3 \right) + V(\mathbf{r}) \right] \Psi_\tau, \label{SWSM} \\
H_2&=&\Psi^\dagger_\tau \left[ \sigma_1  \frac{\partial^2_2-\partial^2_1}{2m} - \sigma_2 \frac{2 \partial_1 \partial_2}{2 m}  -i  v \tau \sigma_3 \partial_z + V(\mathbf{r}) \right] \Psi_\tau, \label{DWSM}  
\end{eqnarray} 
respectively, where $v\sim t a$, $m^{-1} \sim t_1 a^2$, $\tau=\pm$ represent left and right chiral sectors respectively, and we set $t=t'=t_1=1=a$. Effect of random impurities is captured by $V(\mathbf{r})$, distributed uniformly and independently within $[-\frac{W}{2},\frac{W}{2}]$, and the MDOS is numerically evaluated using the kernel polynomial method \cite{KPM-review, herbut-disorder, supplementary}.

\begin{figure}[tbh]
\centering
\includegraphics[width=0.9\columnwidth]{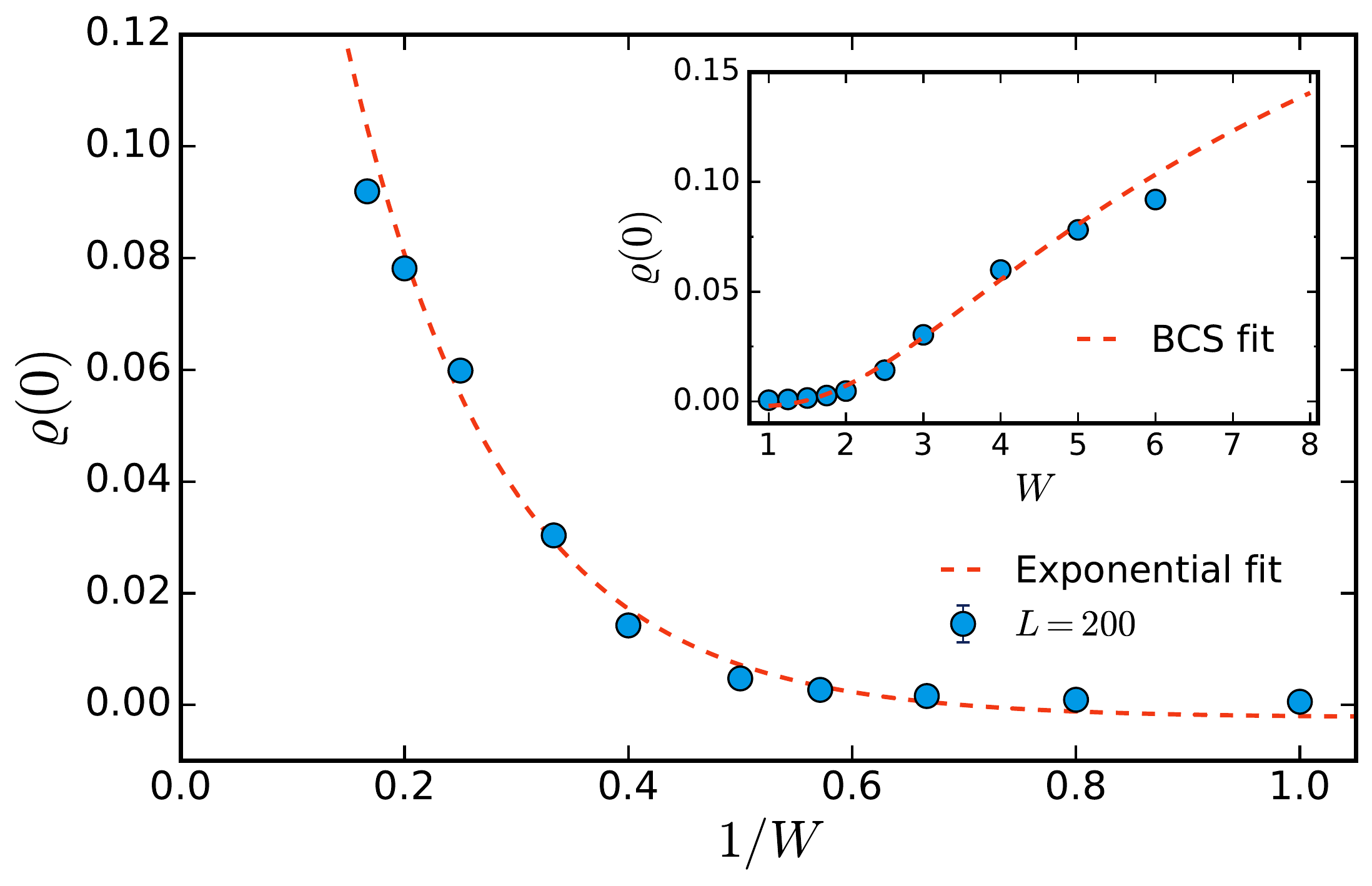}
\caption{(Color online) MDOS at zero energy $\varrho(0)$ in double-WSM as a function of $1/W$. Inset: scaling of $\varrho(0)$ with $W$.}\label{Fig-2}
\end{figure}

To gain insight into the role of disorder in these systems, we can perform disorder averaging, assuming a Gaussian white noise distribution with zero mean, i.e., $\langle \langle V(\mathbf{r}) V(\mathbf{r}') \rangle \rangle= \Delta \delta^3 \left( \mathbf{r}-\mathbf{r}'\right)$ and arrive at the replicated Euclidean action 
\begin{eqnarray}
 \bar{S}_n &=& \int d^3 x dt \: \left( \Psi^\dagger_a \big[ \partial_t + \tilde{H}_n \big]\Psi_a \right)_{(x,t)} \nonumber \\
&-& \frac{\Delta}{2} \int d^{3}x  dt dt' \left(\Psi^\dagger_a  \Psi_a\right)_{(x,t)} (\Psi^\dagger_b  \Psi_b )_{(x,t')},
\end{eqnarray} 
where $a,b$ are replica indices and $\tilde{H}_n$ corresponds to the Hamiltonian from Eqs.~(\ref{SWSM}) and (\ref{DWSM}) in the clean limit. The scale invariance of physical observables ($v$ and $m$) dictates the following space-time(imaginary) 
scaling ansatz: $(x,y) \to e^{l/n} (x,y)$, $z \to e^{l} z$ and $t \to e^{l} t$, accompanied by the 
rescaling of fermionic field $\Psi \to e^{-\left( \frac{1}{n}+ \frac{1}{2} \right) l} \; \Psi$, where $l \sim \log\frac{L}{a}$ is the scaling parameter. The scaling dimension of disorder coupling is $[\Delta]=1 -\frac{2}{n}$. Hence, sufficiently weak disorder is an \emph{irrelevant} (since $[\Delta]=-1$) and a \emph{marginally relevant} (since $[\Delta]=0$) perturbation in WSM and double-WSM, respectively~\cite{supplementary}. Therefore, WSM (double-WSM) is expected to be stable (unstable) in the presence of sufficiently weak randomness.

Notice that $[\Delta] \equiv 2 z-d$~\cite{supplementary}, where $d$ is dimensionality of the system and $z$ is the dynamic critical exponent, together governing the scaling of mean DOS $\varrho(\varepsilon)\sim |\varepsilon|^{d/z-1}$. Therefore, with $z=1$ and $\frac{3}{2}$, $\varrho(\varepsilon)\sim |\varepsilon|^2$ and $|\varepsilon|$, respectively for WSM and double-WSM, in agreement with our numerical findings, see Fig.~\ref{Fig-1}.

\emph{Stability}: WSM evidently remains stable for weak disorder ($W \lesssim  3.3$) and MDOS at zero energy $\varrho(0)=0$, in agreement with our scaling theory [see Fig.~\ref{Fig-1}(a)]. However, for strong disorder WSM appears to undergo a QPT and enter into the CDM phase, where $\varrho(0)$ becomes finite. A finite energy phase diagram of a dirty WSM is shown in Fig.~\ref{Fig-4}(a). This observation is in qualitative agreement with various field theoretic~\cite{fradkin, shindou-murakami, ominato-koshino, goswami-chakravarty, radzihovsky-gurarie-1, roy-dassarma}, and numerical analyses for three dimensional Dirac~\cite{imura, herbut-disorder, pixley-1, pixley-2} and Weyl~\cite{brouwer-2, Ohtsuki, chen-Song} semimetals. We will discuss the nature of such QPT in a moment.

The scaling analysis suggests a BCS-like instability of double-WSM toward the formation of CDM for infinitesimal randomness, and a phase diagram of this system is shown in Fig.~\ref{Fig-4}(b). By contrast, available data of $\varrho(0)$ for double-WSM in a finite system suggest a putative threshold value of disorder ($W_{th} \approx 1$), only beyond which $\varrho(0)$ is visibly finite [Fig.~\ref{Fig-1}(b)]. To examine whether the observed finite-$\varrho(0)$ is a consequence of a $W=0^{+}$ instability, we compare $\varrho(0)$ vs. $1/W$ (see Fig.~\ref{Fig-2}). It is evident that larger systems are required to pin the exponential onset of $\varrho(0)$ for sufficiently weak $W$, as $W_{th} \sim 1/\log(L)$. Still, $\varrho(0)$ depicts a overall good agreement with exponential decrease with $1/W$. In addition, $\varrho(0)$ fits very well with the celebrated BCS scaling form $\varrho(0) \sim \exp(-\lambda/W)$, with non-universal parameter $\lambda=7.2 \pm 0.8$ [see the inset of Fig.~\ref{Fig-2}]. Instability of double-WSM against weak enough disorder is, however, insensitive to the nature of disorder, and similar outcome holds for magnetic disorders~\cite{supplementary}. But, $\varrho(0)$ starts to deviate from BSC scaling for $W>5.0$ and falls below the exponential line. Such behavior can be attributed to the well-known Anderson transition of a three dimensional metal, across which MDOS does not display critical behavior, but decreases monotonically~\cite{pixley-1}. We anticipate that for $W>5.0$ the double-WSM falls within the basin of attraction of metal-Anderson insulator critical point. However, due to a \emph{logarithmic} onset of a metallic phase in dirty double-WSM, for sufficiently weak disorder and/or small system size, quasiparticle excitation can retain their ballistic nature over a large energy scale [see Fig.~\ref{Fig-4}(b)].

\begin{table}
\begin{tabular}{|c|c|c|c|c|c|}
\hline
$N_W$ & $W_c$ & $z$ & $\nu_M$ & $\nu_W$ & $\nu_L$ \\
\hline \hline 
$1 $ & $3.3(0.1)$ & $1.42(0.05)$ & $0.97(0.1)$ & $0.72(0.2)$ & $0.95(0.1)$ \\
\hline
$2 $ & $2.5(0.1)$ & $1.38(0.05)$ & $1.1(0.15)$ & $0.72(0.2)$ & $1.1(0.15)$ \\
\hline
$4 $ & $2.2(0.1)$ & $1.49(0.05)$ & $0.86(0.06)$ & $0.8(0.15)$ & $0.9(0.1)$ \\
\hline
\end{tabular}
\caption{Comparison of critical disorder for WSM-CDM QPT ($W_c$), dynamic critical exponent ($z$), and 
correlation length exponent ($\nu$) extracted from the scaling of MDOS (see text) for WSM with $N_W=1,2,4$~\cite{supplementary}. Quantities in parentheses denote the fitting error. Near WSM-CDM QPT at strong disorder all nodes get coupled and with increasing back-scattering channels or $N_W$, $W_c$ gradually decreases. } \label{table-1} 
\end{table}

\emph{Criticality}: Now we investigate the scaling of MDOS across disorder driven QPT in WSM. Total number of states $\mathcal{N} (\varepsilon, L)$ below the energy $\varepsilon$ in a system of linear size $L$ is proportional to $L^d$, and in general is a function of two dimensionless variables $L/\xi$ and $\varepsilon/\varepsilon_0$. While the correlation length diverges as $\xi\sim \delta^{-\nu}$, the corresponding energy scale vanishes according to $\varepsilon_0 \sim \delta^{\nu z}$, as one approaches the QCP ($\delta \to 0$), where $\delta=\left( W-W_c\right)/W_c$ measures the deviation from the QCP ($W_c$) and $\nu$ is the correlation length exponent~\cite{sachdev, herbut-book}. Consequently,   
\begin{equation}
\mathcal{N} (\varepsilon, L)= \left(L/\xi \right)^{d} \mathcal{G} \left(\varepsilon \delta^{-\nu z}, L^{1/\nu} \delta \right),
\end{equation}
where $\mathcal{G}$ is an unknown scaling function. From the definition of MDOS $\varrho(\varepsilon, L)=L^{-d} d\mathcal{N} (\varepsilon, L)/d\varepsilon$, we then arrive at the following scaling ansatz  
\begin{equation}\label{scaling-gen}
\varrho (\varepsilon, L)= \delta^{\nu(d-z)} \mathcal{F} \left( |\varepsilon| \delta^{-\nu z}, L^{1/\nu} \delta \right),
\end{equation} 
after accounting the particle-hole symmetry, $\varrho(\varepsilon,L)=\varrho(-\varepsilon,L)$, where $\mathcal{F}$ is also an unknown, but universal scaling function. Below we demonstrate the scaling analysis of MDOS in WSM with $N_W=1$ and the results for $N_W=1, 2$ and $4$ are summarized in Table~\ref{table-1}~\cite{supplementary}.

\begin{figure}[htb]
\includegraphics[width=0.99\columnwidth]{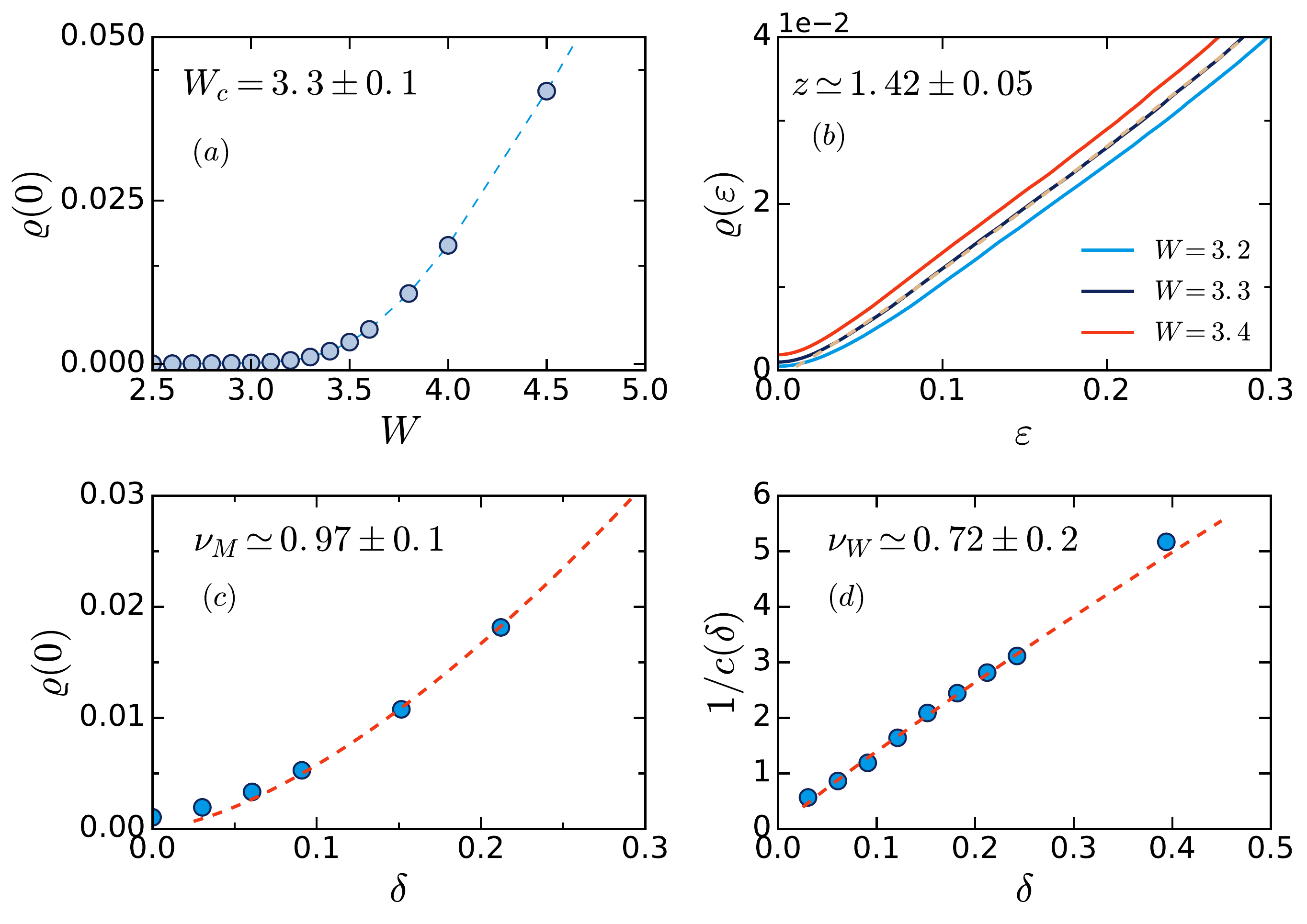}
\caption{(Color online) (a) MDOS at zero energy $\varrho(0)$ vs. disorder; (b) MDOS $\varrho(\varepsilon)$ vs. $\varepsilon$ for $W=3.2, 3.3, 3.4$; (c) $\varrho(0)$ vs. $\delta$, where $\delta=(W-W_c)/W_c$; (d) $c(\delta)^{-1}$ vs. $\delta$, with $c(\delta)=\delta^{(z-1)d \nu}$, for WSM with $N_W=1$, assuming $W_c=3.3$~\cite{supplementary}.}\label{Fig-3}
\end{figure}

First we consider a sufficiently large system ($L=220$), so that the finite size effects are negligible and $L$-dependence in Eq.~(\ref{scaling-gen}) can be ignored. From the scaling of MDOS at zero energy $\varrho(0)$ with disorder we estimate critical disorder for WSM-CDM QPT $W_c=3.3\pm 0.1$ [Fig.~\ref{Fig-3}(a)]. At the QCP ($\delta=0$), the $\delta$-dependence of $\varrho(\varepsilon)$ must cancel out, demanding $F(x) \sim x^{\frac{d}{z}-1}$, and therefore  $\varrho(\varepsilon) \sim |\varepsilon|^{\frac{d}{z}-1}$. From Fig.~\ref{Fig-3}(b) we obtain $z=1.42 \pm 0.05$.  

In the metallic phase, the MDOS at zero energy $\varrho(0)$ is finite and serves as the order parameter. In this regime $\varrho(0) \sim \delta^{(d-z) \nu}$ and one can identify $(d-z) \nu$ as the order parameter exponent ($\beta$). However, such power law dependence of $\varrho(0)$ is valid when $\xi \ll L$. Therefore, we fit $\varrho(0)$ as $\delta^{(d-z) \nu}$ for $\delta \geq 0.06$, and obtain $\nu_M= 0.97 \pm 0.1$, where $\nu_M$ is the correlation length exponent extracted from the metallic phase [see Fig.~\ref{Fig-3}(c)].  

In WSM, the mean DOS scales as $\varrho(\varepsilon) \sim c(\delta)^{-1} |\varepsilon|^{d-1}$, so that we recover $\varrho(\varepsilon) \sim |\varepsilon|^2$ for $d=3$, where $c(\delta) \sim \delta^{(z-1) d \nu_{W}}$ and $\nu_W$ is the correlation length exponent extracted from the WSM phase. However, it should be noted that for $W<W_c$, mean DOS displays a smooth crossover from $|\varepsilon|^2$ (for small $\varepsilon$) to $|\varepsilon|$ (for large $\varepsilon$) dependence. Therefore, estimation of $\nu_{W}$ depends crucially on the range over which we attempt to fit $\varrho(\varepsilon) \sim|\varepsilon|^2$, and accuracy of $\nu_{W}$ can be questioned. Nevertheless, by fitting the coefficient of $|\varepsilon|^2$ with $c(\delta)^{-1}$, we obtain $\nu_{W} =0.72 \pm 0.2$ [see Fig.~\ref{Fig-3}(d)].

\begin{figure}[htb]
\includegraphics[width=1.0\columnwidth]{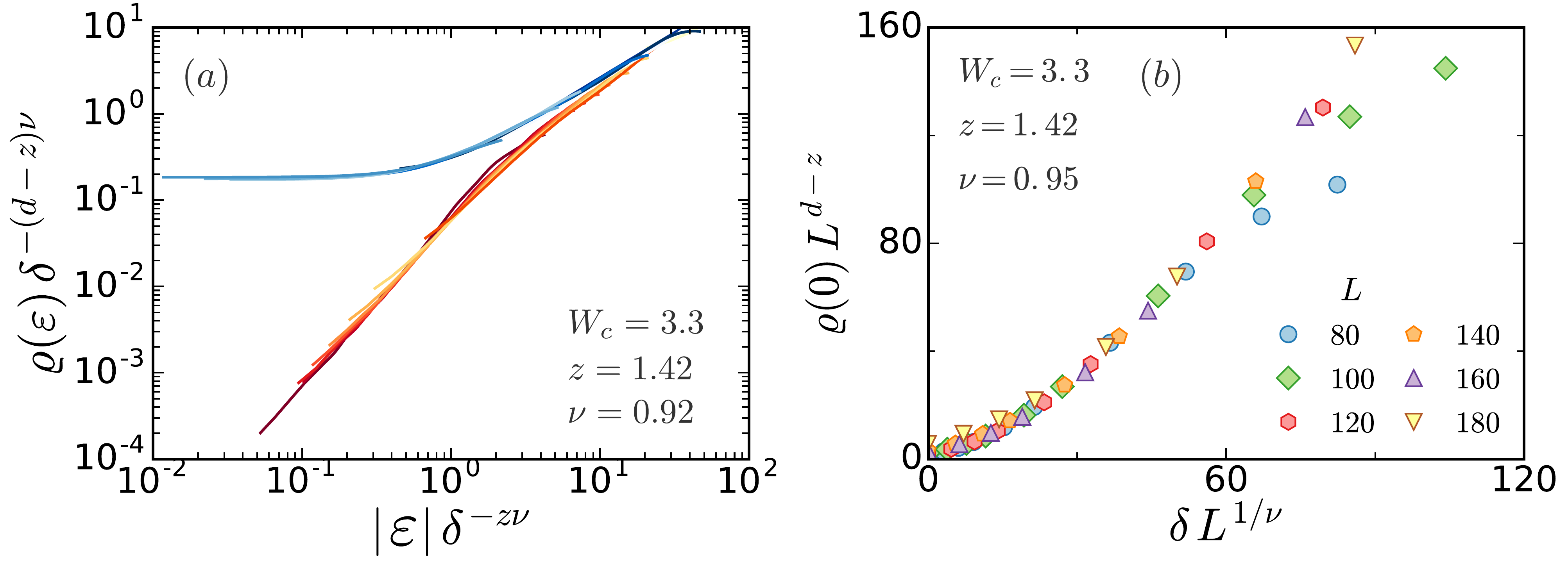}
\caption{(Color online) (a) Single parameter scaling of MDOS in WSM ($N_W=1$) with $L=220$. Top (bottom) branch corresponds to CDM (WSM). (b) Data collapse for $\varrho(0,L)$ when $N_W=1$. For large $\delta$ deviation from single-parameter scaling stem from the Anderson transition at strong disorder.}\label{Fig-5}
\end{figure}

\emph{Data collapse}: We now demosntrate that $\varrho(\varepsilon)$ display a single parameter scaling across WSM-CDM QPT. First, we compare $\varrho(\varepsilon) \delta^{-(d-z)\nu}$ vs. $|\varepsilon| \delta^{-z \nu}$ for $L=220$. Neglecting the high energy part of spectrum ($|\varepsilon|>0.5$) outside the weyl cones and extreme small energy ($|\varepsilon|<10^{-2}$) where numerical accuracy is small, we find that all data from Fig.~\ref{Fig-1}(a) collapse onto two separate branches, associated with the CDM and WSM phases [see Fig.~\ref{Fig-5}(a)]. 

Next we delve into the finite size data collapse for MDOS at $\varepsilon=0$ and estimate $\nu$ independently. Setting $\varepsilon=0$ in Eq.~(\ref{scaling-gen}), we obtain $\varrho(0,L) =L^{z-d} \mathcal{F}(0, \delta L^{1/\nu} )$. An excellent data collapse is achieved by comparing $\varrho(L,0) L^{d-z}$ with $\delta L^{1/\nu}$ for several systems with $80<L<180$, $W_c=3.3$ and $z=1.42$ [see Fig.~\ref{Fig-5}(b)]. The correlation length exponent extracted from the best quality data collapse is $\nu_L=0.95 \pm 0.1$. Thus, $\varrho(0)$ displays a single-parameter scaling and serves as an bonafide order parameter across the WSM-CDM QPT.

\emph{Critical regime}: The crossover from the quadratic (at small energy) to the linear (for higher energies) scaling of $\varrho(\varepsilon)$ allows us to estimate the crossover boundary between the WSM and critical regime at finite energy when $W<W_c$. For sufficiently weak disorder, $\varrho(\varepsilon) \sim |\varepsilon|^2$ over a wide range of energy. As the randomness is gradually increased (but still $W<W_c$), more and more degrees of freedom need to be integrated out (in the spirit of renormalization group) to wash out the effect of disorder from the system. Consequently, the energy window over which $\varrho(\varepsilon) \sim |\varepsilon|^2$ gets reduced and the region where $\varrho(\varepsilon) \sim |\varepsilon|$ increases, with increasing disorder. With this notion we numerically estimate the crossover boundary between the WSM and the critical regime at finite energy [see Fig.~\ref{Fig-4}(a)]. When $W=W_c$, the mean DOS displays a $|\varepsilon|$-linear dependence over the entire energy range ($|\varepsilon|<0.5$). For $W>W_c$, $|\varepsilon|$-linear behavior of $\varrho(\varepsilon)$ ceases at finite energy, defining the boundary between the CDM and critical regime [see Fig.~\ref{Fig-4}(a)].

\emph{Conclusions}: To conclude we show that WSM is stable against weak disorder, but undergoes a QPT and becomes a CDM at strong disorder. Across such QPT MDOS display single-parameter scaling. The critical exponents ($\nu,z$) appear to be independent of the number of Weyl nodes ($N_W$) [see Table~\ref{table-1}], since contribution from any \emph{fermionic-bubble} vanishes in the vanishing replica limit~\cite{roy-dassarma}. The extend of the critical regime at finite energy associated with such QCP [see Fig.~\ref{Fig-4}(a)] can also be measured from ARPES or scaling of specific heat ($C_v \sim T^{d/z}$) in various WSMs~\cite{taas-1, tasas-2, taas-3, nbas-1, tap-1, nbp-1,nbp-2, tas, borisenko, chiorescu} and topological Dirac semimetals (two superimposed copies of WSMs), such as Cd$_2$As$_3$~\cite{cdas}, Na$_3$Bi~\cite{nabi}. In contrast, the double-WSM becomes CDM for weak (infinitesimally small in the thermodynamic limit) disorder. While in the metallic phase $C_v \sim T$, in WSM and double-WSM specific heat scales as $C_v \sim T^3$ and $T^2$, respectively~\cite{supplementary}. Therefore, as a function of temperature specific heat in double-WSM should display a smooth crossover from $T^2$ to $T$-dependence as temperature is gradually decreased [see Fig.~\ref{Fig-4}(b)]. Generalization of scaling analysis dictates that weak disorder is a \emph{relevant} perturbation in \emph{triple-WSM} (monopole charges $\pm 3$), since $[\Delta]=1-\frac{2}{n}=\frac{1}{3}$ for $n=3$. Therefore, among various three-dimensional topological semimetals only conventional WSM is stable against weak disorder~\cite{Goswami-Nevy}. 

 Finally, we discuss the transport phenomena in disordered Weyl systems. In weakly disordered WSM ($W<W_c$) the optical conductivity (in collisionless regime) displays a smooth crossover from $\sigma_{jj}(\Omega) \sim \Omega$ to $\Omega^{1/z}$ dependence as frequency ($\Omega$) is increased, closely following the phase diagram in Fig.~\ref{Fig-4}(a) for $j=x,y,z$~\cite{juricic}. In strong disorder regime ($W>W_c$) $\sigma_{jj}(\Omega)$ becomes finite as $\Omega \to 0$. By contrast, in double- and triple WSMs $\sigma_{zz}(\Omega) \sim \Omega$, while $\sigma_{xx/yy}(\Omega) \sim \Omega^{1/n}$ at high frequency with $n=2$ and $3$, respectively. However, as $\Omega \to 0$, $\sigma_{jj}$ becomes finite in these two systems for arbitrary strength of disorder and for any $j$. Scaling of dc conductivity (collision dominated) follows the ones those for optical conductivity upon taking $\Omega \to T$. Thus, in future one can probe the transport properties to establish the global phase diagram of disordered Weyl materials at finite frequency and temperature.

\emph{Acknowledgements}: J. D. S. and B. R. were supported by the start up grant of J. D. S. from University of Maryland. We thank the visitor program of Max Planck Institute for Complex Systems, Dresden for hospitality during the final stage of the work. We are thankful to J. H. Bardarson, S. Das Sarma, P. Goswami, I. F. Herbut, K. Imura, V. Juri\v{c}i\'{c} and J. Pixley for valuable discussions.

\pagebreak
\onecolumngrid

\begin{center}
{\bf Supplementary Materials for ``\emph{Dirty Weyl semimetals: Stability, phase transition and quantum criticality}"} \\

Soumya Bera$^{1}$, Jay D. Sau$^2$, Bitan Roy$^{2}$ \\

$^1$\emph{Max-Planck-Institut f\"ur Physik Komplexer Systeme, 01187 Dresden, Germany} \\
$^2$\emph{Condensed Matter Theory Center and Joint Quantum Institute, University of Maryland, College Park, Maryland 20742-4111, USA}
 
\end{center}

The Supplementary Materials contain:
\begin{enumerate} 
\item The renormalization group analysis and additional numerical results for disordered double-WSM, 
\item Computation of scaling dimension of disorder coupling for arbitrary quasi-particle dispersion in a $d$-dimensional Weyl system, 
\item The numerical analysis for WSM with $N_W=2$ and $4$, where $N_W$ is the number of Weyl pairs, 
\item Description of tight-binding model for WSM and double-WSM, 
\item Some essential details of numeric methods, which we employed to analyze the behavior of MDOS in dirty WSM as well as double-WSM.
\item Scaling of specific heat ($C_v$) with temperature ($T$) for a dirty double-WSM.   
\end{enumerate}

$\bullet \:$ {\bf \emph{\color{red}RG analysis and additional numerical results for double-WSM:}} While scaling analysis suggest that disorder is a marginal perturbation in double-WSM, to establish that weak disorder is marginally relevant one needs to account for the quantum corrections. The replicated action (imaginary time) in the presence of generic disorder reads as 
\begin{eqnarray}
 \bar{S} &=& \int d^3 x dt \: \Psi^\dagger_a \big[ \partial_t + \sigma_1  \frac{\partial^2_2-\partial^2_1}{2m} - \sigma_2 \frac{2 \partial_1 \partial_2}{2 m}  -i  v \tau \sigma_3 \partial_z \big]\Psi_a {(x,t)}
- \frac{\Delta_0}{2} \int d^{3}x  dt dt' \left(\Psi^\dagger_a  \Psi_a\right)_{(x,t)} (\Psi^\dagger_b  \Psi_b )_{(x,t')} \nonumber \\
&-& \sum^{3}_{j=1} \frac{\Delta_j}{2}\int d^{3}x  dt dt' \left(\Psi^\dagger_a \sigma_i \Psi_a\right)_{(x,t)} (\Psi^\dagger_b \sigma_i \Psi_b )_{(x,t')}, \label{DWSM-action}
\end{eqnarray}
where $a,b$ are replica indices. Here we allowed all types of disorder with distinct bare strengths to examine the low energy behavior in dirty double-WSM in the presence of disorder of arbitrary nature. 
\\

Upon integrating out the fast Fourier modes with $ \Lambda e^{-l} < \sqrt{ \left( \frac{k^2_\perp}{2m} \right)^2 + v^2 k^2_z} <\Lambda $, where $\Lambda$ is the ultraviolet cut-off for energy, we arrive at the following renormalization group (RG) flow equations to quadratic order in disorder coupling 
\begin{eqnarray}
\frac{d \Delta_0}{dl} &=& \Delta^2_0 + \frac{3}{2} \Delta_0 \left( \Delta_1 +\Delta_2 \right) + \Delta_1 \Delta_2+ \Delta_0 \Delta_3, \:\:
\frac{d \Delta_1}{dl} = \frac{1}{2} \left[ \Delta^2_0 + \Delta^2_2 + 2\Delta_1 \Delta_2 + 3 \Delta_1 \Delta_3+\Delta^2_3 +2 \Delta_0 \Delta_2 -\Delta_0 \Delta_1 \right], \nonumber \\
\frac{d \Delta_2}{dl} &=& \frac{1}{2} \left[ \Delta^2_0 + \Delta^2_1 + 2 \Delta_1 \Delta_2 + 3 \Delta_2 \Delta_3 + \Delta^2_3 + 2 \Delta_0 \Delta_1 -\Delta_0 \Delta_2 \right], \: \:
\frac{d \Delta_3}{dl}=\frac{\Delta_3}{2} \left( \Delta_1+\Delta_2 \right), \label{RG-DWSM}
\end{eqnarray}
after taking $\Delta_j m/(2 \pi v) \to \Delta_j$ for $j=0,1,2,3$. Notice that disorder coupling $\Delta_3$ does not get generated through coarse graining if the bare model is deviod of such elastic scatterer. Therefore, a bare model with $\Delta_3=0$, remains closed under RG. However, both $\Delta_1$ and $\Delta_2$ gets generated from $\Delta_0$, thus these three couplings needs to be simultaneously accounted for to keep the model closed under RG. Interestingly if we start with a bare model with only $\Delta_0 \neq 0$, $\Delta_1(l)=\Delta_2(l)$ for all RG time ($l$), and one arrives at a simpler set of flow equations  
\begin{eqnarray}
\frac{d \Delta_0}{dl}= \Delta^2_0 + \Delta^2_\perp +3 \Delta_0 \Delta_\perp, \quad
\frac{d \Delta_\perp}{dl}= \frac{1}{2} \left( \Delta^2_0 + 3 \Delta^2_\perp +\Delta_0 \Delta_\perp \right), 
\end{eqnarray}   
at the one-loop level, where $\Delta_1=\Delta_2=\Delta_\perp$. The above set of flow equations [see Eq.~(\ref{RG-DWSM})] supports only one \emph{unstable} fixed point at $(\Delta_0, \Delta_1, \Delta_2, \Delta_3)=(0,0,0,0)$. Thus weak disorder of arbitrary nature is always a marginally relevant perturbation in double-WSM and drives the system immediately into the CDM phase. 
\\

\begin{figure}[htbp]
\subfigure[]{
\includegraphics[width=7.00cm, height=5.00cm]{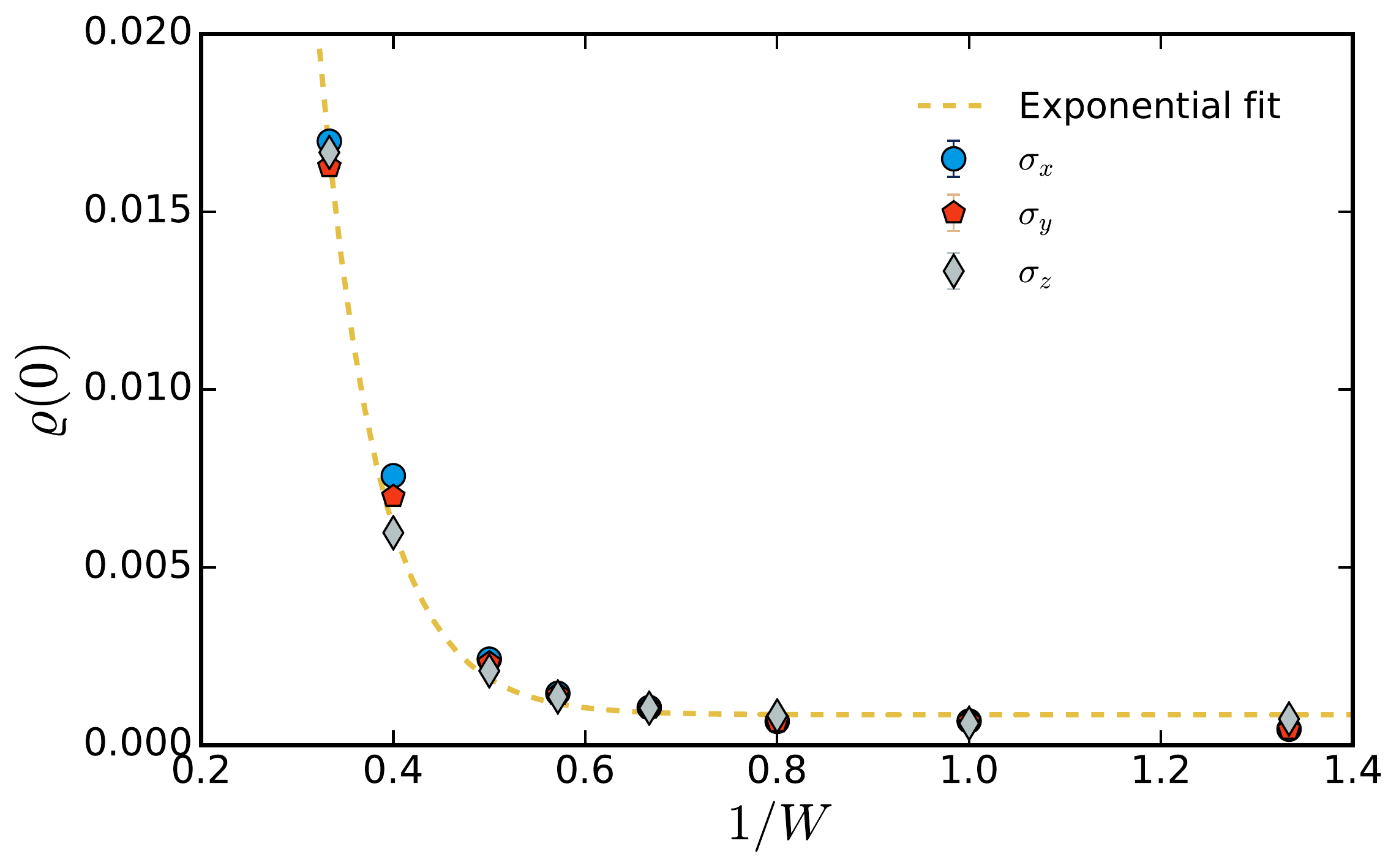}
}
\subfigure[]{
\includegraphics[width=7.00cm, height=5.00cm]{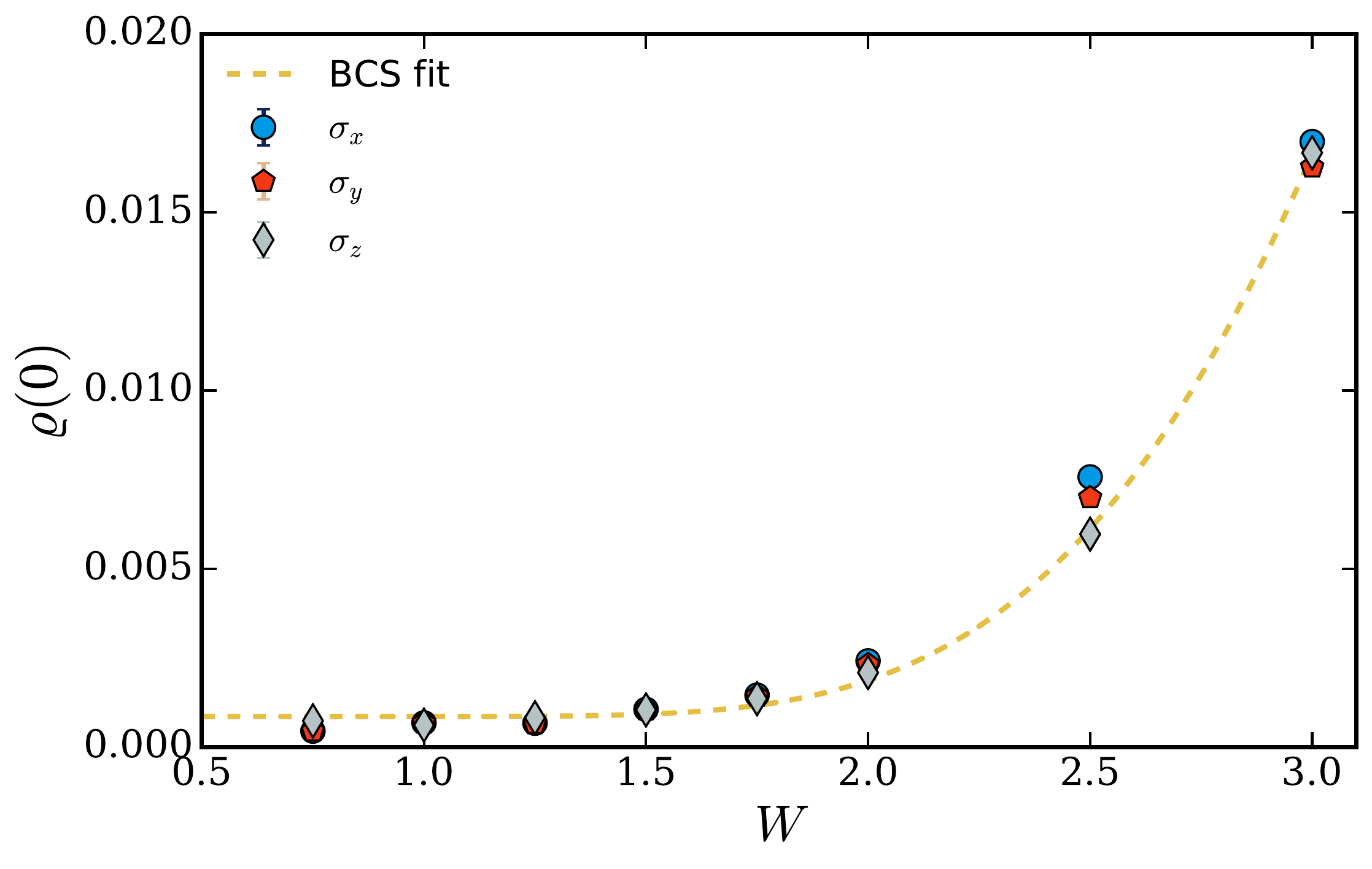}
}
\caption{(Color online) (a) Mean density of states (MDOS) at zero energy in double-Weyl semimetal in the presence of various types of magnetic impurities, denoted by the Pauli matrices [see also Eq.~(\ref{DWSM-action})]. Here, $W$ denotes the strength of disorder. Notice that MDOS at zero energy falls on the same exponential curve irrespective of the nature of impurity scatterer. Here, we used the fitting function $\varrho(0)=a_0 + a_1 \exp \left( -a_2 W^{-1}\right)$, with $a_0=8.6 \times 10^{-4} \pm 9.06 \times 10^{-5}$, $a_1=3.8 \pm 0.76$, $a_2=1.64 \pm 0.59$. (b) BCS scaling of MDOS at zero energy $\varrho(0)$ vs. $W$, $\varrho(0) \sim \exp \left(-\lambda/W \right)$, with $\lambda=1.65 \pm 0.597$.}\label{DWSM-ADOS-Mag}
\end{figure}

\begin{figure}[htbp]
\subfigure[]{
\includegraphics[width=6.5cm, height=5.5cm]{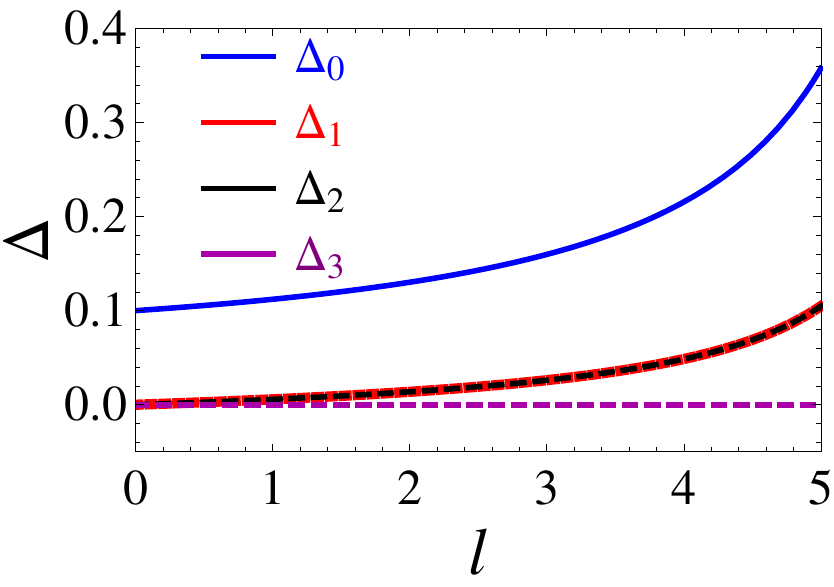}
\label{2weyl}
}\hspace{0.2cm}
\subfigure[]{
\includegraphics[width=6.5cm,height=5.5cm]{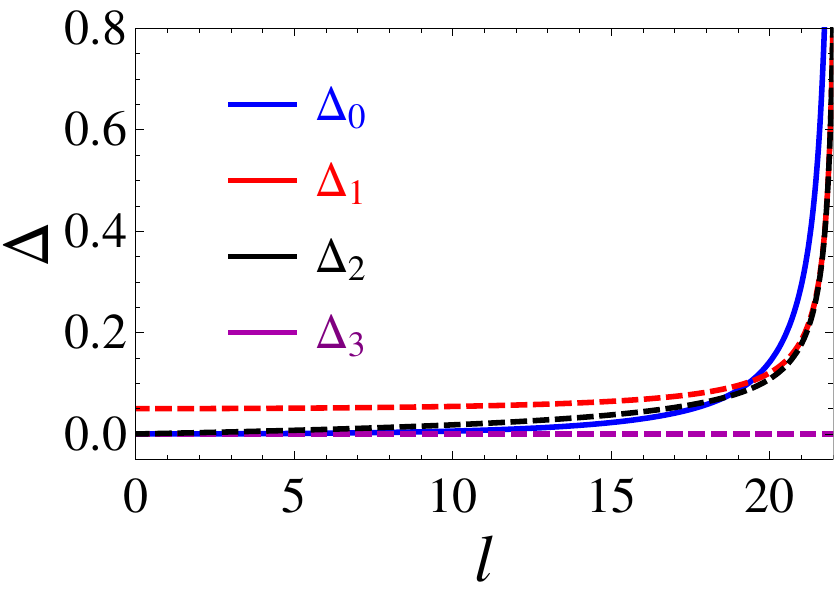}
\label{4weyl}
}
\subfigure[]{
\includegraphics[width=6.5cm,height=5.5cm]{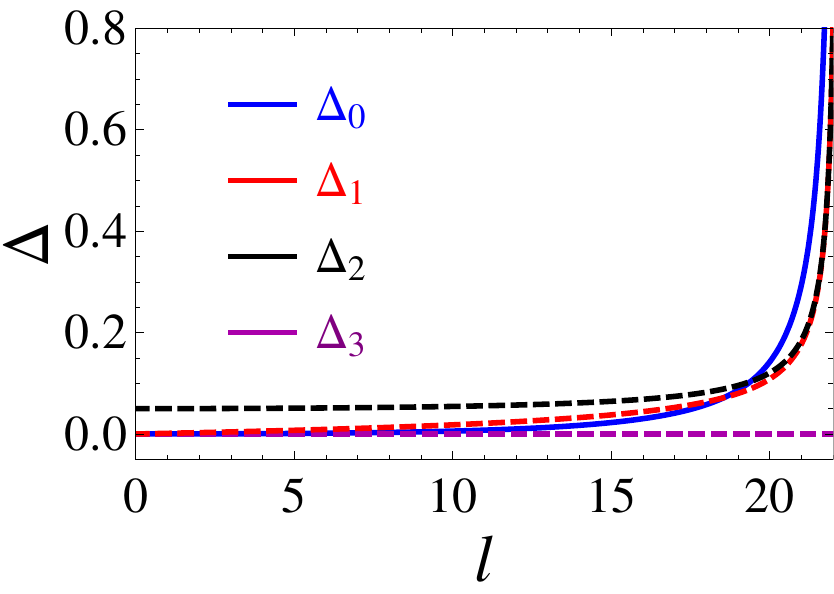}
\label{4weyl}
}\hspace{0.2cm}
\subfigure[]{
\includegraphics[width=6.5cm,height=5.5cm]{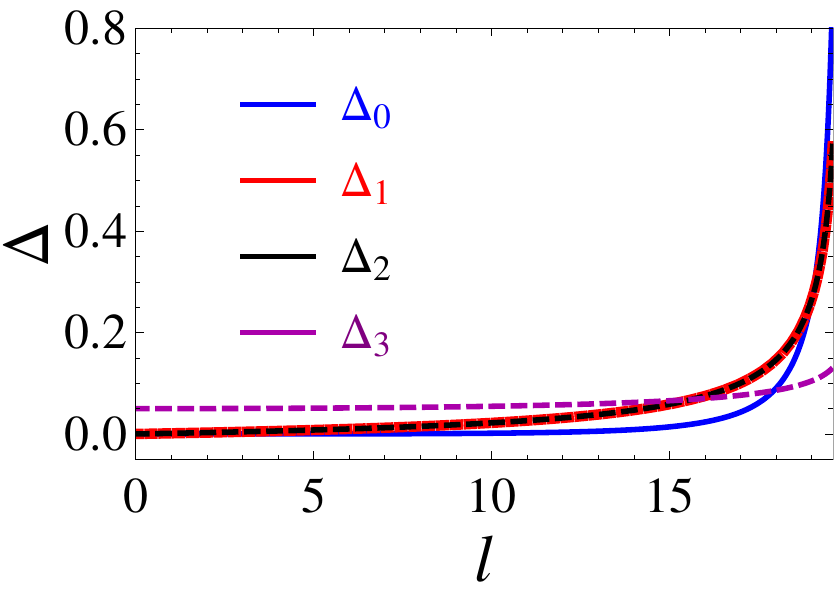}
\label{4weyl}
}
\caption{(Color online) Flow of disorder couplings for (a) $\Delta_0(0)=0.1$, and $\Delta_1(0)=\Delta_2(0)=\Delta_3(0)=0$; (b) $\Delta_1(0)= 0.05$ and $\Delta_0(0)=\Delta_2(0)=\Delta_3(0)=0$; (c) $\Delta_2(0)= 0.05$ and $\Delta_0(0)=\Delta_1(0)=\Delta_3(0)=0$; (d) $\Delta_3(0) =0.05$ and $\Delta_0(0)=\Delta_1(0)=\Delta_2(0)=0$, where $\Delta_j(0)$ represents the bare value of dimensionless disorder coupling (see text). Here, $l$ is the RG time. }\label{RG-flow-DWSM}
\end{figure}

To test the validity of our RG analysis we performed numerical analysis of MDOS at zero energy for all four different types of disorder. Result for regular potential disorder ($\Delta_0$) is shown in the main part of the paper. In this Supplementary Material we present the numerical results for MDOS at zero energy in double-WSM in the presence of three different types of magnetic disorder $\Delta_1$, $\Delta_2$ and $\Delta_3$, and results are shown in Fig.~\ref{DWSM-ADOS-Mag}. The fact that ADOS at zero energy for arbitrary magnetic disorder ($\Delta_j$ for $j=1,2,3$) falls on one curve is quite interesting and can be substantiated from the solution of the RG flow equations, displayed in Eq.~(\ref{RG-DWSM}). As shown in Fig.~\ref{RG-flow-DWSM}, irrespective of the initial conditions, a particular disorder coupling $\Delta_0$ (associated with potential disorder/random charge impurities) always diverges first, leading to BCS-like instability of double-WSM for arbitrarily weak disorder of any nature. Consequently, the MDOS at zero energy for all magnetic disorder falls on the same curve as shown in Fig.~\ref{DWSM-ADOS-Mag}. It is worth mentioning that irrespective of disorder coupling, we always find that $\Delta_1 (l)=\Delta_2(l)$ for $l \gg 1$, as shown in Fig.~\ref{RG-flow-DWSM}.  
\\

To test the rotational symmetry of double-WSM in the $x-y$-plane in the presence of any disorder, we next compute the two-point correlation function or the self-energy correction to the leading order, which is given by 
\begin{equation}
\Sigma(i \omega, {\mathbf k})= (-i \omega) \; \left( \Delta_0 +\Delta_1 +\Delta_2 +\Delta_3 \right) \frac{l}{2}
\end{equation} 
that only depends on external frequency ($\omega$). Thus, rotational symmetry of double-WSM in the $x-y$ plane remains unaffected in the presence of arbitrary disorder, and the system becomes unstable toward a CDM phase for arbitrarily weak disorder. 
\\

Finally, to establish the BCS-scaling of MDOS one needs to find the self-consistent solution of the quasi-particle scattering life-time ($\tau$) at zero energy, since MDOS at zero energy $\varrho(0)$ is inversely proportional to $\tau$~\cite{shindou-murakami}. By employing self-consistent Born approximation we arrive at~\cite{Goswami}
\begin{equation}
\Delta \int_{0}^{\Lambda} d\varepsilon \: \frac{\rho(\varepsilon)}{\hbar^2 \tau^{-2} + \varepsilon^2} =1 \: \Rightarrow \: \varrho(0) \propto \frac{\hbar}{\tau}= \Lambda \exp \left( -\frac{A}{\Delta}\right), 
\end{equation}   
after substituting $\varrho(\varepsilon)\sim |\varepsilon|$, where $A$ is a non-universal model and material dependent constant, and $\Lambda$ is the high-energy cut-off. Thus $\varrho(0)$ displays BCS-scaling as we found in our numerical analysis and predicted from the RG analysis. 
\\

Notice that here we considered only the intra-valley disorder potentials. For Anderson transition one needs to account for the inter-valley scattering as well. In a numerical analysis such inter-valley scattering is always present. However, inclusion of inter-valley scattering does not alter the instability of ballistic fermions in double-WSM toward the formation of a metallic phase for arbitrary weak disorder.   
\\

$\bullet \:$ {\bf \emph{\color{red}Scaling dimension of disorder:}} We now demonstrate the scaling dimension of disorder coupling in a $d$-dimensional system with arbitrary quasi-particle dispersion, captured by the low-energy Hamiltonian $H(\mathbf k)$. The imaginary-time action for such system after performing the disorder averaging is given by 
\begin{eqnarray}
 \bar{S} = \int d^3 x dt \: \left( \Psi^\dagger_a \big[ \partial_t + H(\mathbf k \rightarrow -i \nabla) \big]\Psi_a \right)_{(x,t)}
- \frac{\Delta}{2} \int d^{3}x  dt dt' \left(\Psi^\dagger_a  \Psi_a\right)_{(x,t)} (\Psi^\dagger_b  \Psi_b )_{(x,t')},
\end{eqnarray} 
 where $a$ and $b$ are replica indices. We here perform a slightly different rescaling of space-time(imaginary) coordinates from the one shown in the main part of the paper, under which $\left( \mathbf x, t\right) \to \left( e^{l}\mathbf x, e^{z l} t\right)$ and the parameter $z$ is chosen such that all physical parameters in $H(\mathbf k \to -i \nabla)$, such as Fermi velocity ($v$), quasi-particle mass ($m$), remain scale invariant. The parameter $z$ also determines the scaling of MDOS according to $\varrho(\varepsilon) \sim |\varepsilon|^{d/z-1}$. The replicated action $\bar{S}$ then remains scale invariant when the fermionic field is rescaled according to $\Psi \to e^{-dl/2}\Psi$. Under such rescaling of coordinates and field, the disorder coupling $\Delta \to e^{(d-2z)l}\Delta$. Therefore, scaling dimension of disorder coupling is $[\Delta]=2z-d$. 
\\

From the main part of the paper, where we performed an anisotropic scaling over various spatial coordinates (to keep $v$ and $m$ scale invariant) with $t \to e^{l} t$, we know that $[\Delta]=-1$ for WSM and $0$ for double-WSM. Therefore, for these two systems $z=1$ and $3/2$, respectively. As a result, the MDOS in these two systems respectively scales as $\varrho(\varepsilon) \sim |\varepsilon|^2$ and $|\varepsilon|$. These results are announced in the main part of the paper and also consitent with numerical computation of MDOS in WSM and double-WSM, shown in Fig.~1 of the main part of the paper.    
\\

\begin{figure}[htb]
\subfigure[]{
\includegraphics[width=8.00cm, height=6.50cm]{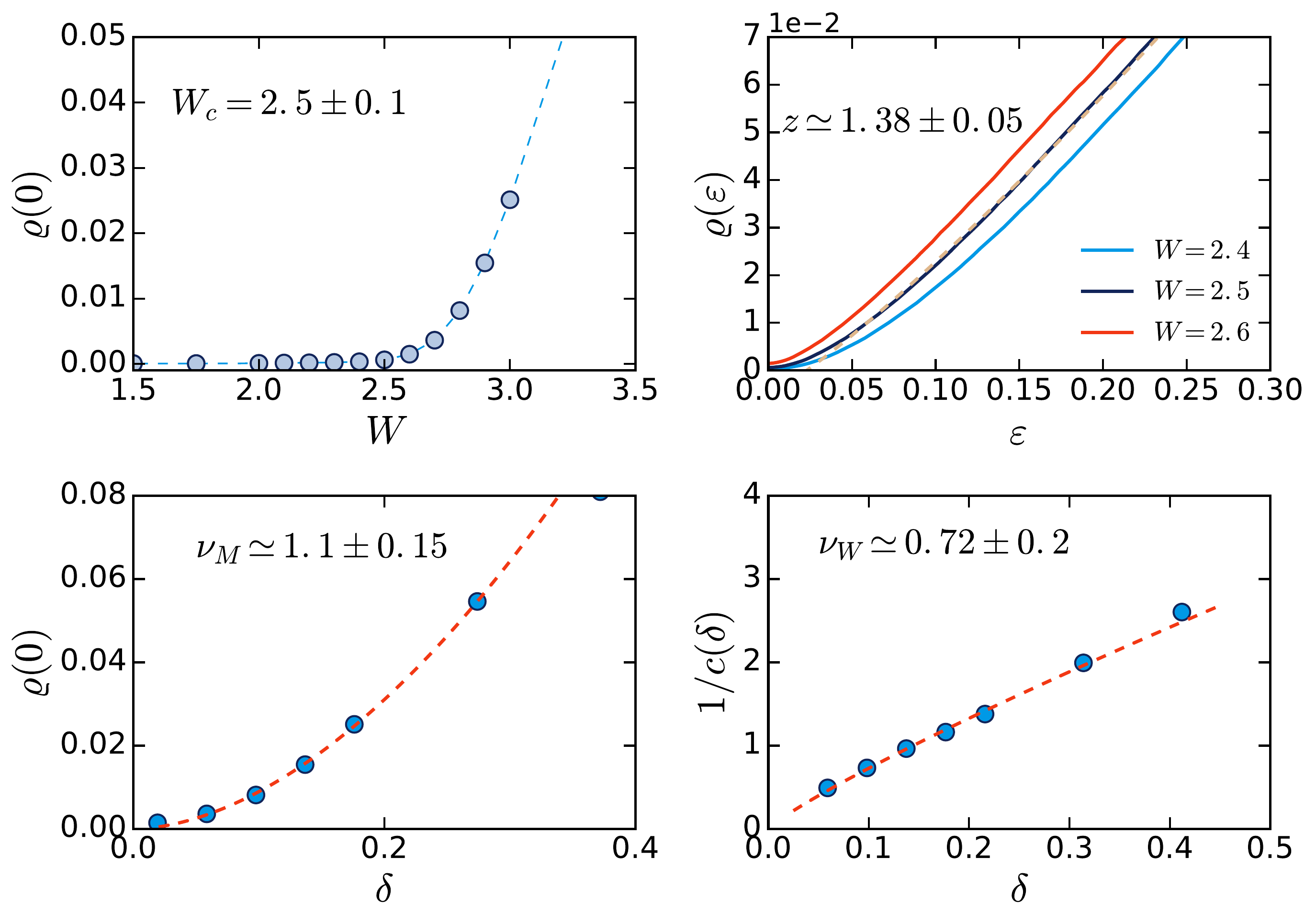}
\label{2weyl}
}
\subfigure[]{
\includegraphics[width=8.00cm, height=6.50cm]{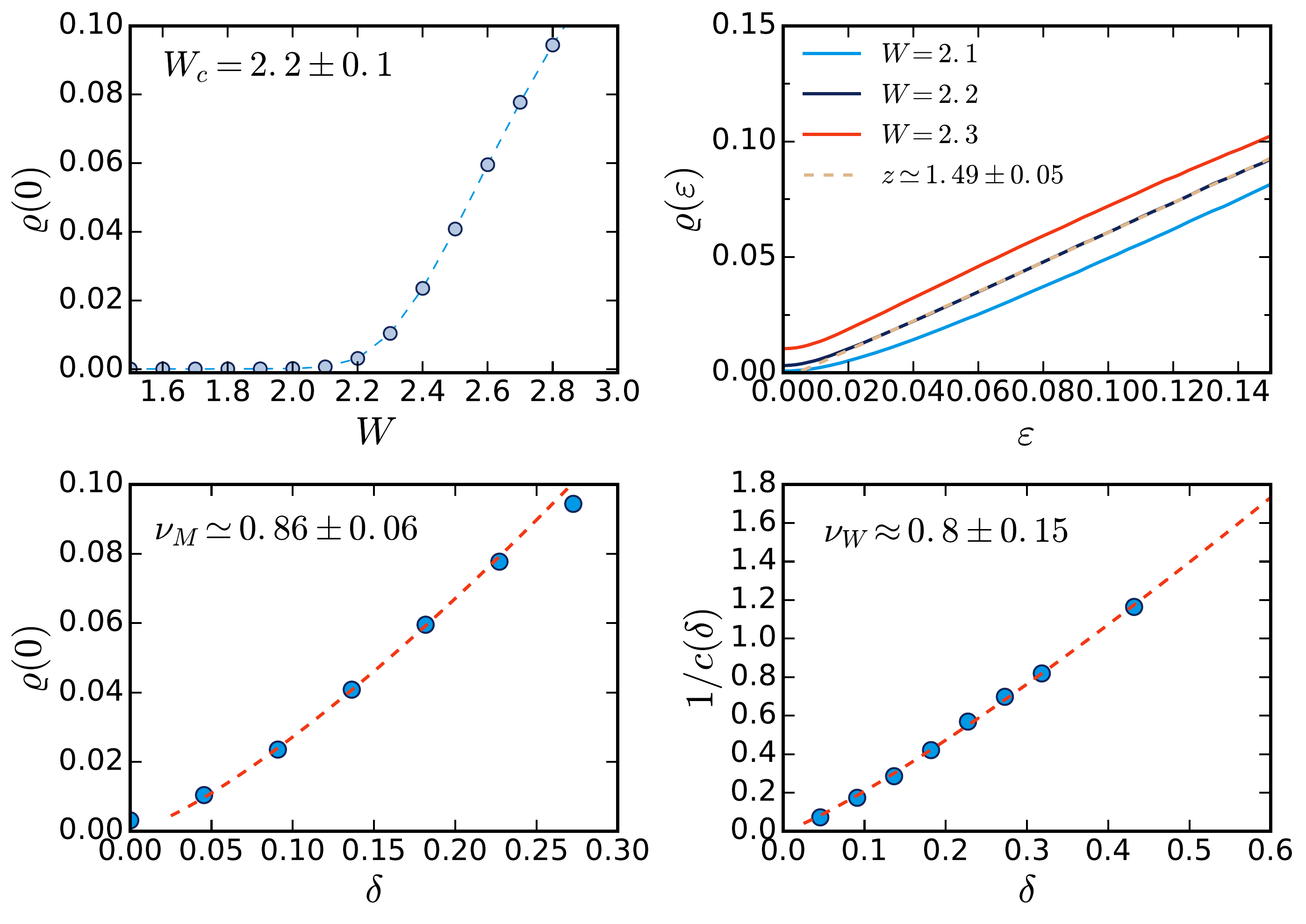}
\label{4weyl}
}
\caption{(Color online) (a) Mean DOS at zero energy $\varrho(0)$ vs. disorder ($W$), mean DOS $\varrho(\varepsilon)$ vs. $\varepsilon$ for $W=2.4, 2.5, 2.6$, $\varrho(0)$ vs. $\delta$, where $\delta=(W-W_c)/W_c$, and $c(\delta)^{-1}$ vs. $\delta$, where $c(\delta)=\delta^{(z-1)d \nu}$, for WSM with $N_W=2$ (from left to right, and top to bottom). (b) Similar quantities for $N_W=4$. Extracted values of $W_c$, $z$ and $\nu$ are quoted in the figures.} \label{analysis}
\end{figure}

\begin{figure}[htb]
\subfigure[]{
\includegraphics[width=8.50cm, height=3.75cm]{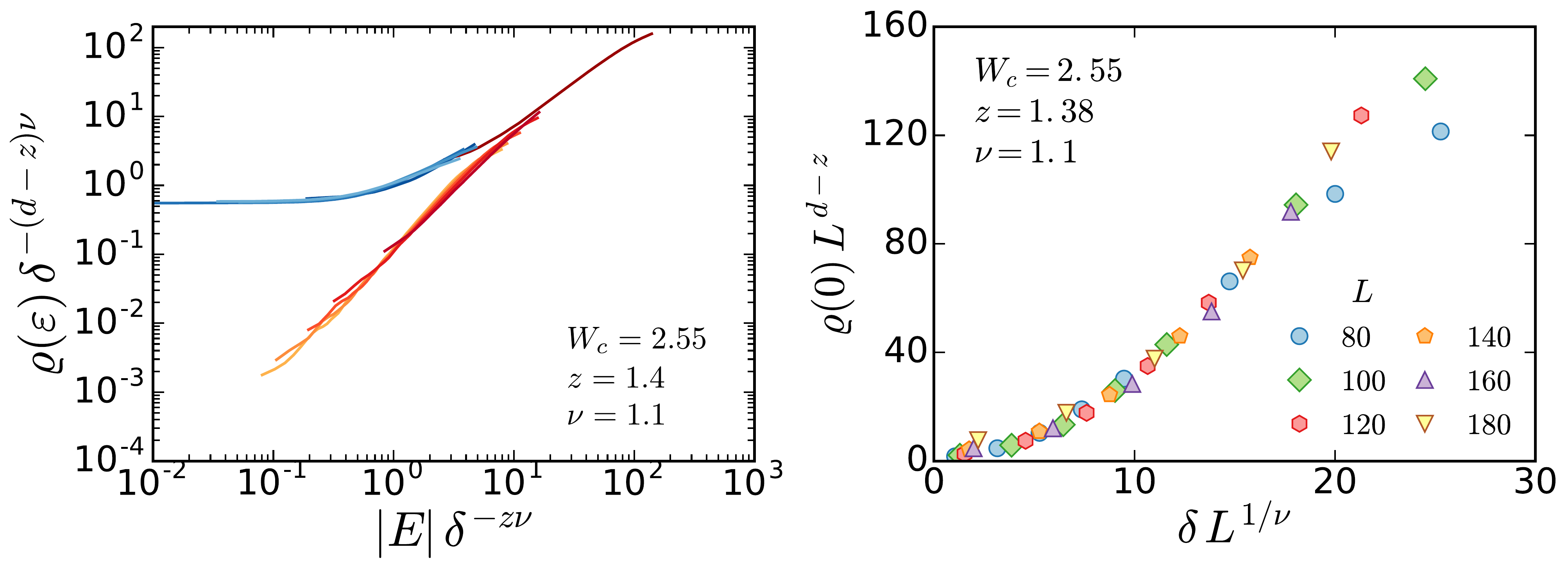}
\label{2weyl}
}
\subfigure[]{
\includegraphics[width=8.50cm, height=3.75cm]{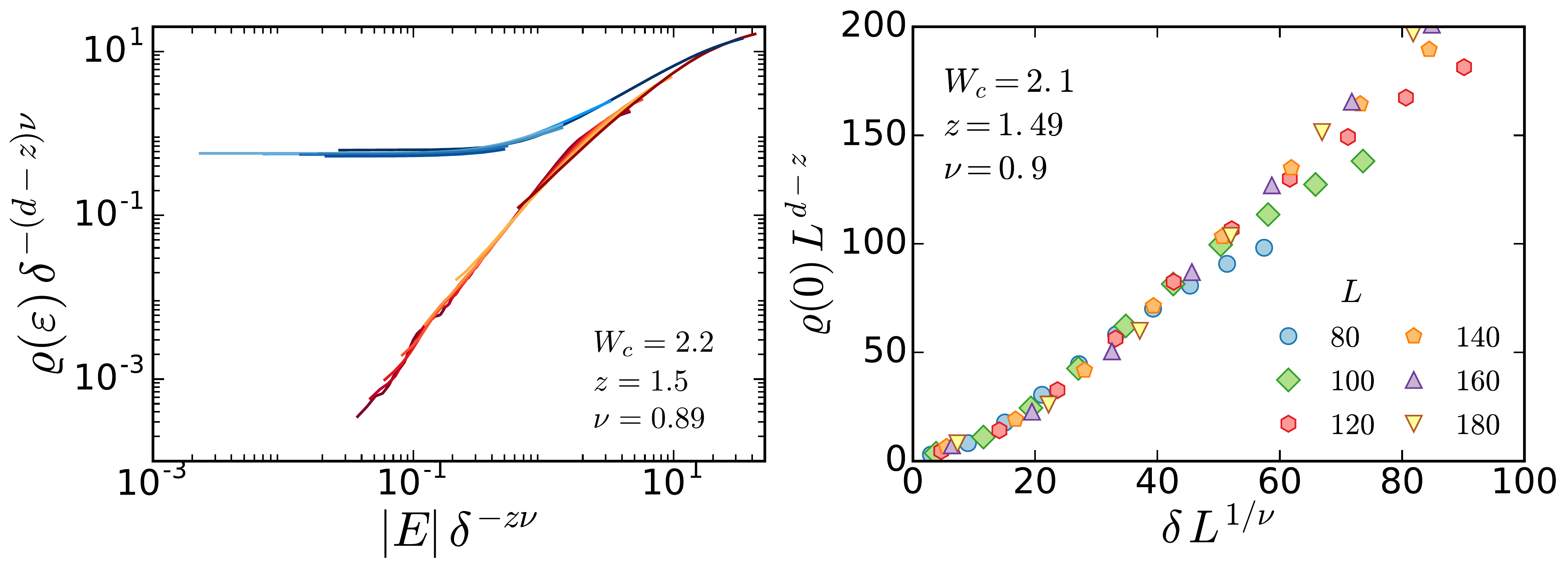}
\label{4weyl}
}
\caption{ (a) Single-parameter scaling of MDOS at finite (left) and zero (right) energy in WSM with $N_W=2$. (b) Similar quantities for $N_W=4$. Top (bottom) branch in the left panel of (a) and (b) corresponds to metal (Weyl semimetal). }\label{collapse-24}
\end{figure}

$\bullet \:$ {\bf \emph{\color{red}Numerical analysis for WSM with $N_W=2$ and $4$:}} We now present the numerical analysis for $W_c$ (critical disorder for WSM-CDM transition), and computation of various critical exponents ($\nu$ and $z$) in WSM with $N_W=2$ and $4$, where $N_W$ is the number of Weyl pairs. The results are already quoted in Table~I of main part of the paper. These quantities are extracted by using identical methods, described in the main paper for $N_W=1$. The corresponding analysis for $N_W=2$ and $4$ are presented in Fig.~\ref{analysis}. The data collapse for MDOS at finite energies in WSM and CDM phases, as well as that for MDOS at zero energy in the CDM phase for $N_W=2$ and $4$ are shown in Fig.~\ref{collapse-24}.  
\\

\begin{figure}[htbp]
\includegraphics[width=7.00cm, height=5.50cm]{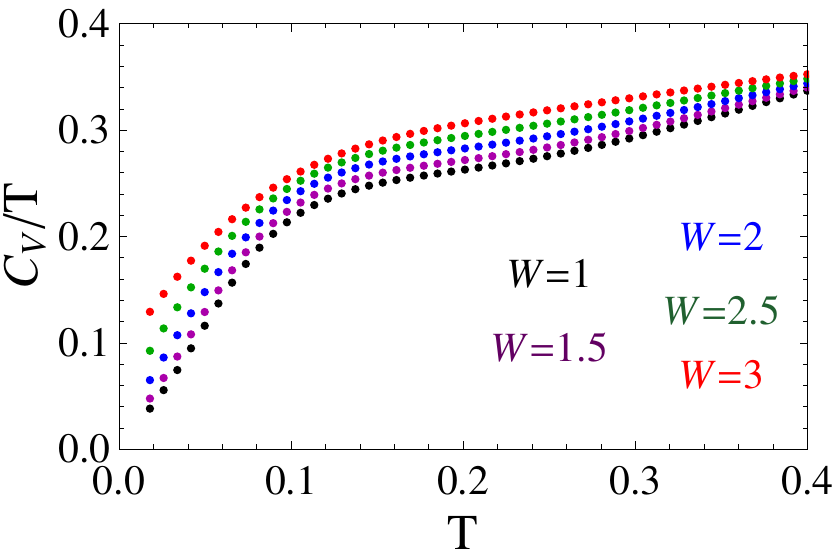}
\caption{(Color online) Scaling of $C_v/T$ with temperature $T$ (measured in units of $t$) in dirty double-WSM. Strength of disorder is quoted in the figure.  }\label{specificheat}
\end{figure}

$\bullet \:$ {\bf \emph{\color{red} Tight-binding model for WSM and  double-WSM:}} The tight-binding model that support a WSM, is described by the following momentum dependent form factors, $N_1(\mathbf{k})=t \sin(k_1 a)$, $N_2(\mathbf{k})=t \sin(k_2 a)$, $N^{1}_3(\mathbf{k})=t\cos(k_3 a)/2$ and $N^{2}_3(\mathbf{k})= t'[b-\cos(k_1 a) -\cos(k_2 a)]$, where $a$ is the lattice spacing and $N_3(\mathbf{k})= N^{1}_3(\mathbf{k}) + N^{2}_3(\mathbf{k})$. The Hamiltonian is then given by
\begin{equation}
H_W= \sum_{\mathbf{k}} \Psi^\dagger_{\mathbf{k}} \left[ N_1(\mathbf{k}) \sigma_1 + N_2(\mathbf{k}) \sigma_2 + N_3(\mathbf{k}) \sigma_3 \right] \Psi_{\mathbf{k}},
\end{equation}
which on a cubic lattice with periodic boundary in each direction translates into the follwoing tight-binding model 
\begin{eqnarray}
H = \sum_{\mathbf{r}} \bigg[ \frac{t}{4}\; \Psi^\dagger_{\mathbf{r}} \sigma_3 \Psi_{\mathbf{r}+\hat{e}_3} + \sum_{j=1,2} \Psi^\dagger_{\mathbf{r}}  \left[ \frac{i t}{2} \sigma_j -\frac{t'}{2} \sigma_3 \right] \Psi_{\mathbf{r}+\hat{e}_j} 
+ H.c  \bigg] + \sum_{\mathbf{r}}  \Psi^\dagger_{\mathbf{r}} \left[ 2 t' \sigma_3 + V(\mathbf{r}) \right] \Psi_{\mathbf{r}},
\end{eqnarray}
where $\Psi^\top_{\mathbf{r}}=\left(c_{\mathbf{r}, \uparrow}, c_{\mathbf{r}, \downarrow} \right)$, and $c_{\mathbf{r}, s}$ is electronic annihilation operator at site $\mathbf{r}$ with spin projection $s=\uparrow, \downarrow$. Nearest-neighbor sites are connected by the unit vectors $\hat{e}_j$, for $j=1,2,3$. For $b=2$, a pair of Weyl nodes are located at $\mathbf{k}=(0,0,\pm 1) \frac{\pi}{2 a}$, while for $b=0$, two pairs of Weyl nodes are found at $\mathbf{k}=(1,-1,\pm 1)\frac{\pi}{2 a}$ and $(-1,1,\pm 1)\frac{\pi}{2 a}$. Finally, when $t^\prime =0$, four pairs of Weyl nodes are found at $\mathbf{k}=(0,0,\pm 1)\frac{\pi}{2 a}$, $(1,-1,\pm 1)\frac{\pi}{2 a}$, $(-1,1,\pm 1)\frac{\pi}{2 a}$ and $(-1,-1,\pm 1)\frac{\pi}{2 a}$. Therefore, by tuning various parameters of the above tight-binding model one can realize WSMs with different $N_W$. 
\\

As announced in the main part of the paper that by setting $N_1(\mathbf{k})=t_1 [\sin(k_1 a) -\sin(k_2 a)]$, $N_2(\mathbf{k})=t_1 \cos(k_1 a) \cos(k_2 a)$, $N^{1}_3(\mathbf{k})=t\cos(k_3 a)$ and $N^{2}_3(\mathbf{k})= t'[2-\sin(k_1 a) -\sin(k_2 a)]$, one can realize a double-WSM in the vicinity of $\mathbf{k}=(\frac{\pi}{2 a}, \frac{\pi}{2 a}, \pm \frac{\pi}{2 a})$. In cubic lattice with periodic boundary in each direction, the corresponding tight-binding model is    
\begin{eqnarray}
H &=& \sum_{\mathbf{r}} \bigg[ \frac{t}{2} \Psi^\dagger_{\mathbf{r}} \sigma_3 \Psi_{\mathbf{r}+\hat{e}_3} 
+ \frac{i t}{2} \left[ \Psi^\dagger_{\mathbf{r}} \sigma_1 \Psi_{\mathbf{r}+ \hat{e}_1} - \Psi^\dagger_{\mathbf{r}} \sigma_1 \Psi_{\mathbf{r}+ \hat{e}_2} \right] 
+ \frac{t}{4} \sum_{\alpha=\pm} \Psi^\dagger_{\mathbf{r}+ \hat{e}_1} \sigma_2 \Psi^\dagger_{\mathbf{r}+ \alpha \hat{e}_2} -\frac{t'}{2} \sum_{j=1,2} \Psi^\dagger_{\mathbf{r}} \sigma_3 \Psi_{\mathbf{r}+ \hat{e}_j} + H.c  \bigg] \nonumber \\
&+&  \sum_{\mathbf{r}}  \Psi^\dagger_{\mathbf{r}} \left[ 2 t' \sigma_z + V(\mathbf{r}) \right] \Psi_{\mathbf{r}}.
\end{eqnarray}
Here, $V(\mathbf{r})$ captures the effect of random quenched charge impurities. 
\\

$\bullet \:$ {\bf \emph{\color{red}Numerical method:}} We address the stability of various members of the Weyl family by analyzing the MDOS at zero energy. The MDOS is numerically calculated using the kernel polynomial method~\cite{KPM-review, herbut-disorder}. The MDOS is evaluated using the following definition 
\begin{equation}
\varrho(\varepsilon)= \text{Tr} \: \: \delta(\varepsilon-\hat{H}),
\end{equation} 
where Tr is evaluated stochastically and the delta-function is exapnded using Chebyshev polynomial. We usually take $4096$ Chebyshev moments and few ($\sim 8$) trace vectors to calculate MDOS. We ensure the convergence of MDOS with respect to number of Chebyshev moments and number of random vectors. We typically take $\sim 30$ disorder realization to further minimize the residual statistical error, which is quite small in large system~($L=220$) due to the self-averaging nature of MDOS.
\\

$\bullet \:$ {\bf \emph{\color{red} Specific heat ($C_v$) in dirty double-WSM:}} The specific heat is related to the MDOS according to 
\begin{equation}
C_v=\beta^2 \; \int^{\infty}_{-\infty} d\varepsilon \; \frac{\varrho(\varepsilon) \; \varepsilon^2}{4 \; \cosh^2 (\beta \varepsilon/2) },
\end{equation}
where $\beta=1/(k_B T)$ is the inverse temperature and we set $k_B$. In numerical calculation of $C_v$ from the tight-bonding model, integral over energy is restricted by the band width. 
\\

Since in clean double-WSM MDOS $\varrho(\varepsilon) \sim|\varepsilon|$, the specific heat $C_v \sim T^2$. On the other hand, in the metallic phase $C_v \sim T$. To detect the onset of an metallic phase, we thus compare $C_v/T$ vs. $T$ for a wide range of temperature. The results are displayed in Fig.~\ref{specificheat}. For any strength of disorder ($W$) $C_v/T$ becomes finite as $T \to 0$. For weak enough disorder $C_v/T \sim T$ for a wide range of temperature, while such range of $T$ gradually decreases with increasing disorder in the system. Therefore, in the presence of even infinitesimal disorder a double-WSM turns into a diffusive metal at lowest energy scale. However, over a wide temperature regime quasi-particles retains their ballistic nature as reflected from $C_v/T \sim T$ behavior, which, however, diminishes with increasing strength of disorder. 
\\

From Fig.~\ref{specificheat} we also note that with increasing strength of disorder the magnitude of specific heat increases monotonically, stemming from the fact that the MDOS in the system with increasing randomness. Therefore, with increasing disorder system becomes more metallic.

\end{document}